\documentclass{article}

\usepackage{microtype}
\usepackage{graphicx}
\usepackage{subfigure}
\usepackage{booktabs} 

\usepackage{hyperref}

\usepackage[accepted]{icml2023}

\usepackage{amsmath}
\usepackage{amssymb}
\usepackage{mathtools}
\usepackage{amsthm}

\usepackage[capitalize,noabbrev]{cleveref}

\theoremstyle{plain}

\theoremstyle{definition}

\theoremstyle{remark}

\usepackage[disable,textsize=tiny]{todonotes}

\icmltitlerunning{HLSTransform: Energy-Efficient Llama 2 Inference on FPGAs Via High Level Synthesis}

\begin{document}

\twocolumn[
\icmltitle{HLSTransform: Energy-Efficient Llama 2 Inference on FPGAs Via High Level Synthesis}







\icmlsetsymbol{equal}{*}

\begin{icmlauthorlist}
\icmlauthor{Andy He}{equal,corn}
\icmlauthor{Darren Key}{equal,corn}
\icmlauthor{Mason Bulling}{equal,corn}
\icmlauthor{Andrew Chang}{equal,corn}
\icmlauthor{Skyler Shapiro}{equal,corn}
\icmlauthor{Everett Lee}{corn}
\end{icmlauthorlist}

\icmlaffiliation{corn}{Cornell University, Ithaca, NY}

\icmlcorrespondingauthor{Darren Key}{dyk34@cornell.edu}

\icmlkeywords{FPGA, High Level Synthesis, LLM, Llama, Hardware Acceleration}

\vskip 0.3in
]



\makeatletter\def\Hy@Warning#1{}\makeatother
\printAffiliationsAndNotice{\icmlEqualContribution}

\begin{abstract}
Graphics Processing Units (GPUs) have become the leading hardware accelerator for deep learning applications and are used widely in training and inference of transformers; transformers have achieved state-of-the-art performance in many areas of machine learning and are especially used in most modern Large Language Models (LLMs). However, GPUs require large amounts of energy, which poses environmental concerns, demands high operational costs, and causes GPUs to be unsuitable for edge computing. We develop an accelerator for transformers, namely, Llama 2, an open-source state-of-the-art LLM, using high level synthesis (HLS) on Field Programmable Gate Arrays (FPGAs). HLS allows us to rapidly prototype FPGA designs without writing code at the register-transfer level (RTL). We name our method HLSTransform, and the FPGA designs we synthesize with HLS achieve up to a 12.75x reduction and 8.25x reduction in energy used per token on the Xilinx Virtex UltraScale+ VU9P FPGA compared to an Intel Xeon Broadwell E5-2686 v4 CPU and NVIDIA RTX 3090 GPU respectively, while increasing inference speeds by up to 2.46x compared to CPU and maintaining 0.53x the speed of an RTX 3090 GPU despite the GPU’s 4 times higher base clock rate. With the lack of existing open-source FPGA accelerators for transformers, we open-source our code and document our steps for synthesis. We hope this work will serve as a step in democratizing the use of FPGAs in transformer inference and inspire research into energy-efficient inference methods as a whole. The code can be found on \href{https://github.com/HLSTransform/submission}{GitHub}.
\end{abstract}

\section{Introduction}
\label{submission}
Hardware accelerators have long appeared in computing \cite{nvidia-accelerated} to improve performance compared to general-purpose CPUs through specialized operations, high parallelism, and efficient memory systems \cite{domain-specific-hardware}. The use of accelerators for deep learning have been especially significant to accommodate models that are rapidly scaling up in size and complexity, such as transformer-based Large Language Models (LLMs) which have become increasingly complex with a massive influx of research following the advent of OpenAI's ChatGPT. Meta's popular Llama 2 model, for instance, is trained on 2 trillion tokens and ranges up to 70 billion parameters \cite{llama2}. GPUs are currently the dominant accelerators for general deep learning tasks as they can be easily leveraged to develop extremely efficient implementations of parallel basic linear algebra subroutines (BLAS), which are commonly used in deep learning algorithms. \cite{performance-comparison-blas}.

However, the most glaring tradeoff to using GPUs is their massive demand for power, resulting in high carbon emissions and energy costs. The carbon footprint of training Llama 2 is officially estimated at 539 tons carbon dioxide equivalent \cite{touvron2023llama}, which is almost 72x the amount the average US household produces per year at 7.5 tons \cite{forestpreserves}. However, while model training takes large amounts of energy, energy spent running inference on the model is typically larger; NVIDIA and Amazon estimate that over 80\% of their energy usage for AI models is spent in inference, and for Google, 60\% of their energy usage for AI models is for inference \cite{mcdonald2022great} \cite{googlecarbonfootprint}. Inference dominates emissions in ChatGPT-like services from the querying of millions of users, producing 25x the carbon emissions of GPT-3 \cite{samsi2023words} \cite{chien2023reducing}. 

High energy consumption also poses a problem for operational costs and for edge computing applications. High energy consumption forces the inference of LLMs and deep learning models to be mostly allocated to GPU clusters. An article from Sequoia Capital estimates that for data centers, the price from energy alone is roughly equal to the amount spent on buying GPUs \cite{sequoia}. For applications requiring real-time inference on the edge, in addition to monetary reasons, a dedicated GPU is often impractical as it cannot draw sufficient and sustained power.

While GPU acceleration will likely remain dominant in the near future despite the power disadvantage, there is value in exploring different avenues of hardware acceleration as deep learning tasks continue to diverge into highly specific applications. Further, as transformers become more and more ubiquitous, there is a case to be made for designing model-specific hardware accelerators solely to optimize inference. To that end, Field Programmable Gate Arrays (FPGAs) are another desirable choice for accelerators as they offer a hardware reconfigurable for specific tasks enabled by a large number of programmable logic gates, making them inexpensive to iterate hardware designs on. Furthermore, FPGAs are distinguished for their reduced power consumption, which on average is only 28\% of GPUs \cite{cong2018understanding}. 

What limits the adoption of FPGAs currently is the high barrier of entry and relative lack of research compared to GPUs. FPGAs are commonly used to prototype hardware designs for system-on-chip (SoC) and Application Specific Integrated Circuit (ASIC), which is typically done on the register-transfer level (RTL) using hardware description languages like Verilog. However, the design and verification of RTL modules are known to be extremely complex and time-consuming. High Level Synthesis (HLS) is a methodology that seeks to address that complexity by allowing developers to write hardware descriptions in more accessible, high-level languages like C or C++. HLS tools convert high-level code input into RTL code that optimizes for performance, area, and energy consumption, leading to faster prototyping and iteration for FPGAs. Furthermore, the nature of HLS tools and availability of Vitis C / RTL co-simulation make it simple to verify the correctness of the synthesized hardware designs; these factors allow HLS to significantly shorten the traditional hardware development cycle.

In this literature, we employ HLS tools to design FPGAs for accelerating Llama 2 inference. In addition to the large GPU power footprint of LLMs that may be addressed with FPGAs, the complex data flow of transformer models \cite{ftrans} often comprises of nonlinearities or token encoding subroutines (such as RoPE) that are difficult to accelerate on GPUs but could be better suited for FPGAs. Llama 2 is chosen in particular due to its open-source implementations and superb performance \cite{touvron2023llama}, making it a popular and well researched choice. We use Andrej Karpathy's llama2.c repository \cite{llama2c} to develop our methods on a relatively small (110M parameters) model to allow for our financial and compute constraints. We focus on inference over training due to its higher energy usage and greater suitability for FPGAs.

In summary, through our methods which we name HLSTransform, we demonstrate the following:
\begin{enumerate}
    \item \textbf{Low power and energy consumption}

    Energy savings up to a 12.75x reduction of total energy consumption compared to CPU and an 8.25x reduction of total energy consumption compared to GPU.
    
    \item \textbf{Fast inference speeds and low latency}
    
    Acceleration up to 2.46x in inference speed in comparison to CPU, and maintaining up to 0.53x in inference speed in comparison to GPU, despite the GPU having 4x higher base clock rate.

    \item \textbf{Verification of HLS tools for faster deployment}

    Ensuring HLS tools run properly to synthesize appropriate FPGA designs for this study. We also test the learning curve of the tools for the use of any developer without extensive hardware backgrounds.

\end{enumerate}

We open-source our code and document our FPGA synthesis to the public, available in our GitHub repo here: \texttt{\href{https://github.com/HLSTransform/submission}{github.com/HLSTransform/submission}}. To the best of our knowledge, our model is one of the first open-source HLS-based implementations for transformers. In our research process, the lack of documentation for many steps of the process combined with the absence of existing open-source FPGA accelerators for transformers served as a high barrier to entry, and we hope our work serves a step forward in democratizing the usage and research of FPGAs for transformer inference.

\section{Related Work}
We delineate a few studies that relate to FPGA accelerators for transformers and the application of high level synthesis.

\subsection{Existing Hardware Accelerators for Transformers on FPGA}
Existing hardware accelerators for transformers on FPGA incorporate specialized techniques to optimize performance on FPGAs. Column Balanced Block Pruning \cite{blockpruning} and FTrans \cite{ftrans} are two novel frameworks for transformer models suitable for FPGA acceleration. By incorporating weight pruning to employ sparse matrix multiplication, these papers are able to achieve multiple folds of improvements in transformer inference compared to CPUs and GPUs in terms of performance and energy efficiency. We instead strive to maintain dense matrix multiplication in our methods to allow for general application to existing transformer models. Similarly, NPE \cite{npe} introduces a framework for FPGA acceleration on transformers, utilizing piecewise linear approximations for nonlinear functions (e.g. softmax and GELU) to achieve speedups. In contrast, we compute exact values for nonlinear functions. Our methodology allows us to avoid needing to train FPGA-specific models and avoid potential accuracy tradeoffs associated with these novel pruning or approximation techniques. The only potential accuracy tradeoffs are from our usage of quantization, where we follow the well-tested quantization algorithm “Q8\_0”, explored further in Section 3.2.

\subsection {hls4ml}
We aim to inspire the democratization of FPGA accelerators for deep learning using HLS. Fast Machine Learning Lab's hls4ml \cite{hls4ml} is an open-source workflow that enables fast prototyping of machine learning algorithms via Python for FPGAs \cite{software_hls4ml} \cite{Duarte:2018ite}. hls4ml has been successful in being one of the first open source HLS tools for deep learning especially in Python, but a major limitation is its lack of support for attention layers used in transformer models. The tool mainly supports convolutional neural networks (CNNs) and feed-forward deep neural networks (DNNs), but transformer models like Llama 2 are unique in requiring attention layers and novel techniques such as  Rotary Position Embeddings, which are not yet supported by this framework.

\section{Methods}
We follow the same architecture outlined in the original Llama 2 paper \cite{llama2}: the standard Transformer architecture \cite{attention}, rotary position embeddings \cite{rope}, grouped query attention \cite{group}, RMS norm for pre-normalization \cite{rmsnorm}, and the SwiGlu activation function \cite{swiglu}. Since FPGAs are constrained in performance by the amount of on-chip memory, we selected a small 110M parameter model trained on the TinyStories dataset to test our designs \cite{tinystories}. We discuss the limitations of the small model size further in the Limitations and Future Works section. More details on model architecture are included in the Appendix.

\subsection{Implementation}

\begin{figure}[ht]
\vskip 0.2in
\begin{center}
\centerline{\includegraphics[width=\columnwidth]{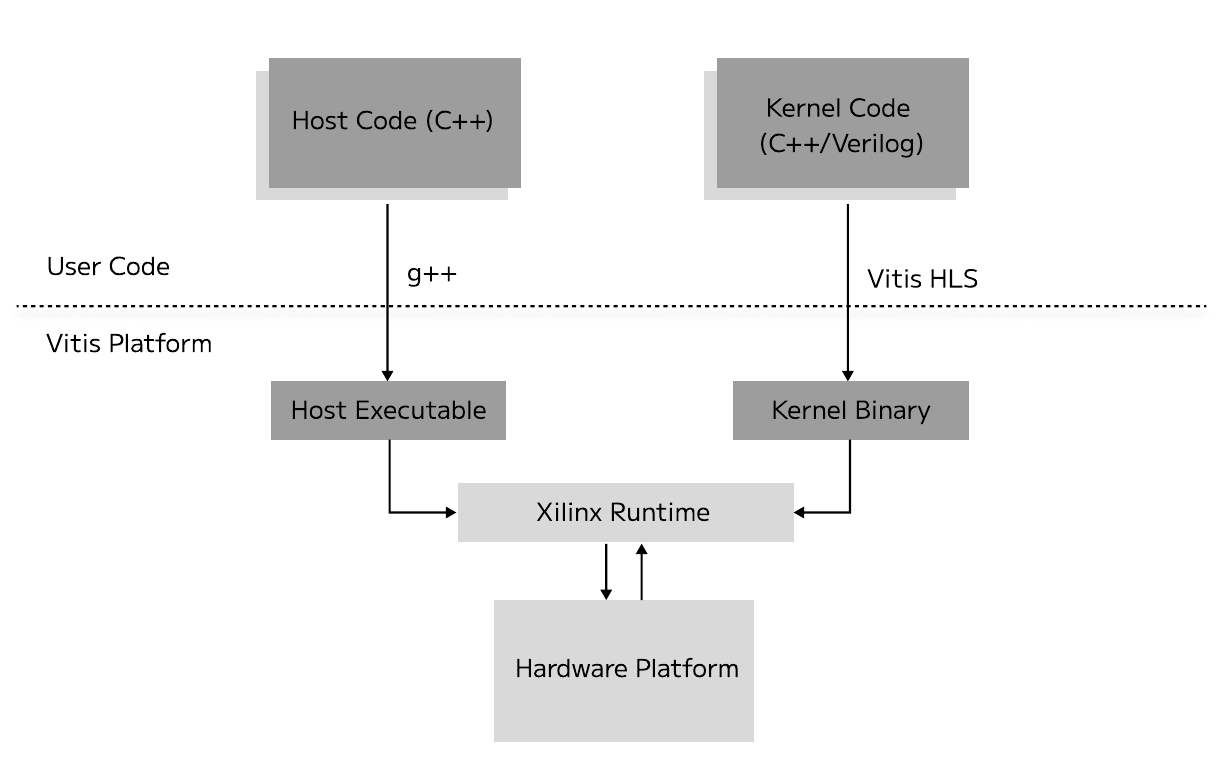}}
\caption{Vitis HLS development workflow.}
\label{figure1}
\end{center}
\vskip -0.2in
\end{figure}

Our implementation of Llama 2 is built on Andrej Karpathy’s llama2.c repository. For our HLS toolchain, we chose Vitis, as it is both widely used and directly supported by the FPGAs available to us on AWS. The code is split into two portions, the host and the kernel. The kernel code contains the hardware description for one iteration of the computationally-intensive forward inference pass and is synthesized for the FPGA, while the host is responsible for driving the kernel code. The host interfaces with the FPGA accelerator through the Xilinx Runtime Library (XRT).

The host sends the input parameters, such as the token and position to the FPGA via direct memory access (DMA). The FPGA is responsible for writing the output to a shared-buffer that can be accessed by both the host and the kernel. The host reads the output and performs sampling to extract the next token. 

We focus on three HLS optimizations: pipelining, unrolling, and array partitioning. We also implement software-level optimizations; in addition to memory limitations, FPGAs also have constraints regarding Digital Signal Processor (DSP) blocks, which are specialized hardware modules within an FPGA that are optimized for efficient  floating point arithmetic calculations. However, the number of available DSP blocks is limited and varies depending on the FPGA model; to address DSP and on-chip memory bottlenecks, we first quantized the weights from 32-bit (single-precision) IEEE floating points to 8-bit signed integers.

\subsection{Int-8 Quantization}
Included in Karpathy's work, we employ an 8-bit integer quantized forward pass to run our inference on FPGAs \cite{llama2c}. The quantization process is post-training; i.e. it is independent from model training.

We perform symmetric quantization, scaling each weight between [-127, 127]. Each weight is divided into sections of equal size, each of which is quantized by the following formula, where $w$ here represents a vector of weights in that section and the square brackets denote the rounding function.

\begin{equation*}
w = \lceil 127 * \frac{w}{\lVert w \rVert_{\infty}} \rfloor
\end{equation*}

This quantization has been noted to perform well empirically, used in Georgi Gerganov’s popular GGML library for efficient CPU transformer inference and referred to as “Q8\_0” quantization in the library \cite{ggml}. We quantize the embedding, attention, and the feedforward weights. The RMSNorm params, which are sensitive to error, are kept in float32 precision.

Although quantization leads to decreased model accuracy, the accuracy dropoff is minimal, and we explore the effects of quantization in Section 4.1. Quantization allows for smaller weights, which permits us to better utilize the limited memory bandwidth on the FPGA and perform integer-only calculations, which provides inference speedups through lower precision arithmetic calculations \cite{ibert}.

\subsection{Optimization of Llama 2 Accelerator Using HLS Pragmas}

Pragmas in High-Level Synthesis (HLS) are directives used to guide the HLS compiler in the process of converting the high-level code into a hardware description, typically used when indicating to the compiler that a specific optimization should be performed on some section of the code.

\begin{figure}[h]
\vskip 0.2in
\begin{center}
\centerline{\includegraphics[width=\columnwidth]{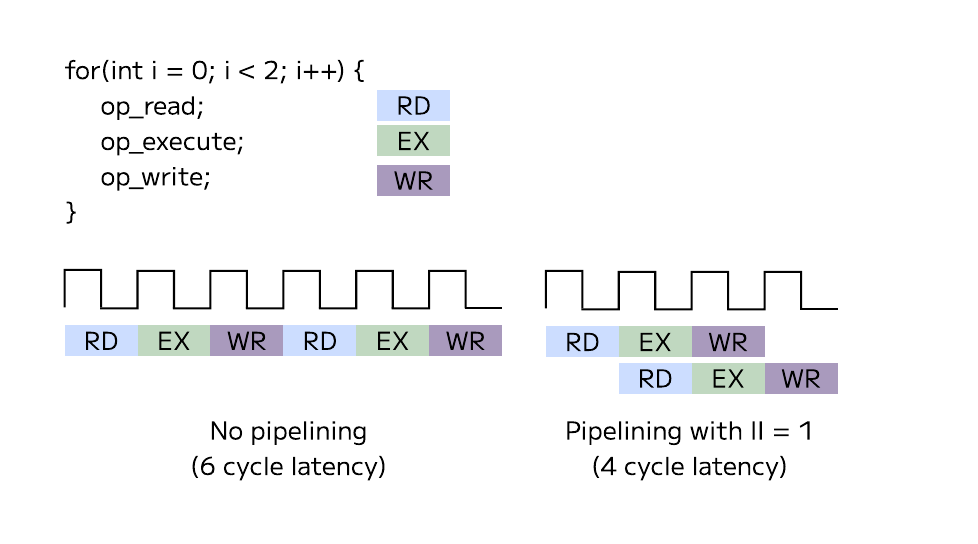}}
\caption{Pipelining two iterations of instructions with read, execute, and write stages.}
\label{figure1}
\end{center}
\vskip -0.2in
\end{figure}

\subsubsection{Pipelining}
Pipelining HLS is a technique used to enhance the performance of hardware circuits generated from high-level code. This method involves dividing a process into several stages, each separated by registers. Similar to an assembly line, pipelining allows different stages of a computation to occur in parallel but on different sets of data. In HLS, this means translating high-level programming constructs into pipelined hardware structures.

For example, in a computation involving multiple arithmetic operations, HLS can break down these operations into stages, where each stage performs a part of the computation. By doing so, while one stage is processing one set of data, the next stage can work on another, leading to increased throughput. The pipeline pragma is applied to the main loops responsible for computing matrix-vector multiplication and rotary position embeddings. 

\subsubsection{Loop Unrolling}
Loop unrolling is an optimization technique that increases the efficiency of hardware implementations derived from high-level code. This process involves expanding the loop body multiple times in order to reduce the number of iterations. By doing this, loop unrolling enables the simultaneous execution of multiple consecutive loop iterations, as long as there are no intra-loop data dependencies. 

In other words, if a loop is executed $N$ times and we unroll it $M$ times, the loop body will be replicated $M$ times within each iteration, thereby reducing the total number of iterations to $N / M$. This technique is especially useful in hardware design because it can lead to more parallel operations, allowing the hardware to perform more tasks simultaneously at the cost of chip space. 

\subsubsection{Memory Partitioning}
The application of HLS partitioning pragmas is a critical step in the design of the Llama 2 deep learning accelerator. Typically, FPGA BRAM is implemented as a dual-port memory, which greatly restricts the degree to which code can be parallelized on chip. By dividing arrays and memory structures into smaller, independent blocks, different data segments can be processed in parallel. Memory partitioning ensures more efficient utilization of the available computational resources, thereby enhancing the throughput for matrix multiplication operations, a common bottleneck in neural network computations.

\subsubsection{Burst Reads / Writes over AXI4}
In general, a dual-port memory bank can support two reads per cycle. Since global memory cannot be partitioned completely due to the limitation on the number of memory channels available to the FPGA, we instead utilize burst reads and writes into local on-chip buffers. By using a technique called widening, global memory can be accessed through dual-port 256-bit wide lines, allowing the simultaneous read of 64 8-bit integers per cycle. Efficient data transfer between the FPGA and external memory is essential, given the large amount of parameters that need to be read from memory before any computations can begin.

\section{Results and Discussion}

We evaluate the perplexity, latency, power, and energy consumption of the 110M parameter Llama 2 model across CPU, GPU, and FPGA. We provide more details of the evaluation setup in the Appendix. We run our benchmarks for 256 tokens and the max context length of 1024 tokens to test both the short and long text generation domains.

 Our FPGA designs were synthesized targeting the Ultrascale+ VU9P platform available on AWS, and the synthesized designs were then exported to an Amazon Machine Image (AMI) using a custom toolchain provided by Amazon \cite{aws-fpga}. We use the f1.2xlarge instance from AWS to host the FPGA, and we use the t2.2xlarge instance for our CPU benchmarks (8 vCPUs, 2.3 GHz Intel Xeon Broadwell E5-2686 v4), the same CPUs used in the FPGA instance, and an NVIDIA RTX 3090 GPU for our GPU benchmarks. We use the original Llama 2 implementation provided by Meta for our GPU experiments. We run all samples with non-batched inference (batch size 1). 

While we run benchmarks of FPGA performance against CPUs and GPUs, we are unable to provide equitable quantized benchmarks for GPUs, as the different scaling factors per section in the quantization algorithm used would require specialized kernels to make this efficient. To provide equitable comparisons, we also provide perplexity benchmarks, a common metric for model quality, along with inference latency and energy consumption benchmarks to demonstrate minimal tradeoffs to accuracy while fully utilizing the optimized integer-arithmetic abilities of FPGAs.

\subsection{Perplexity}

We measure perplexity on the validation dataset for TinyStories, for both the quantized and unquantized models of the 110M parameter model; perplexity is a common metric for model quality that measures a model's uncertainty about its predictions. Our experimental setup is detailed further in the Appendix.

\begin{table} [!h]
\begin{center}
\begin{small}
\begin{sc}
\label{perplexity}
\caption{Perplexity (lower is better)}
\vskip 0.15in
\begin{tabular}{lr}
\toprule
 Model  & Average perplexity (ppl) $\downarrow$ \\ 
\midrule
 Quantized 110M & 2.9679\\ 
 Unquantized 110M & 2.9667 \\ 
 Unquantized 42M & 3.1810\\ 
\bottomrule
\end{tabular}
\end{sc}
\end{small}
\end{center}
\vskip -0.1in
\end{table}

The quantized model is able to retain nearly identical levels of performance (0.04\% increase in perplexity) as the unquantized model while utilizing integer only computations. We include the perplexity benchmark for a 42 million parameter model as reference, which is 7.22\% higher than the unquantized 110 million parameter model.

\subsection{Latency and Speed}

We measure inference latency in milliseconds and inference speed in tokens per second. Similar to NPE, an existing hardware accelerator for FPGAs, we obtain our timing results from the system simulations \cite{npe}, and we provide a report of our full timings in the Appendix.

\begin{table}[h!]
\begin{center}
\begin{small}
\begin{sc}
\caption{Inference speed (tokens per second)}
\label{Speed}
\vskip 0.15in
\begin{tabular}{lcr}
\toprule
 Hardware  & 256 tokens $\uparrow$ & 1024 tokens $\uparrow$ \\ 
\midrule
 CPU & 23.21 toks/s & 19.63 toks/s\\ 
 GPU & 107.00 toks/s & 107.24 toks/s\\ 
 FPGA & 57.11 toks/s & 57.11 toks/s \\
\bottomrule
\end{tabular}
\end{sc}
\end{small}
\end{center}
\end{table}

\begin{table}[h!]
\begin{center}
\begin{small}
\begin{sc}
\caption{Inference Latency (milliseconds)}
\label{Latency}
\vskip 0.15in
\begin{tabular}{lrr}
\toprule
 Hardware  & 256 tokens $\downarrow$ & 1024 tokens $\downarrow$\\ 
\midrule
 CPU  & 43.08 ms & 50.94 ms\\ 
 GPU  & 9.34 ms & 9.32 ms\\ 
 FPGA & 17.51 ms & 17.51 ms \\
\bottomrule
\end{tabular}
\end{sc}
\end{small}
\end{center}
\end{table}

According to Table 2, the FPGA is 2.46x the inference speed of CPU and 0.53x the inference speed of GPU.

Although the GPU performs inference faster than the FPGA, one of the primary bottlenecks of deep learning inference is memory bandwidth and the availability of on-chip memory \cite{onchip}. A RTX 3090 has 24GB VRAM running at 1219 MHz with a base core clock of 1395 MHz \cite{rtx}. In comparison, a VU9P FPGA has 345.9 MB of combined on-chip BRAM and URAM, running at a much slower clock speed of around 200-300 MHz depending on the module; however, with much lower clock speeds, the FPGA is able to achieve better efficiency on power and energy consumption, as shown below.

\subsection{Energy and Power Consumption}
We utilize the CodeCarbon library, also used by HuggingFace to provide carbon estimations for the BLOOM LLM, to provide energy consumption metrics for CPU and GPU performance \cite{mittechreview} \cite{bloom} \cite{codecarbon}. For GPU benchmarks, CodeCarbon sources energy consumption directly from NVIDIA’s NVML library. For the AWS CPU benchmarks, energy consumption cannot be directly sourced since AWS uses hypervisors, and CodeCarbon uses an estimation derived from empirical energy consumption data \cite{codecarbon}. 

As CodeCarbon does not handle FPGA energy consumption measurement, energy consumption metrics for FPGA is provided by Vivado and AWS provided tools \cite{aws-fpga}.

\begin{table}[!h]
\begin{center}
\begin{small}
\begin{sc}
\caption{Power consumption on FPGA (Watts)}
\label{power-fpga}
\vskip 0.15in
\begin{tabular}{lcr}
\toprule
 FPGA  & 256 tokens $\downarrow$ & 1024 tokens $\downarrow$\\ 
\midrule
 Average & 9 W & 9 W \\ 
 Max & 12 W & 11 W \\ 
\bottomrule
\end{tabular}
\vskip -0.1in
\end{sc}
\end{small}
\end{center}
\end{table}

\begin{table} [!h]

\begin{center}
\begin{small}
\begin{sc}
\caption{Average power consumption (Watts)}
\label{avg-power-consumption}
\vskip 0.15in
\begin{tabular}{lrr}
\toprule
 Hardware  & 256 tokens $\downarrow$  & 1024 tokens $\downarrow$  \\ 
\midrule
 CPU & 42.5 W & 42.5 W\\ 
 GPU & 126.9 W & 130.6 W \\ 
 FPGA & 9 W & 9 W \\
\bottomrule
\end{tabular}
\end{sc}
\end{small}
\end{center}

\end{table}
The average power consumption of the FPGA is considerably lower than the average power consumption for both CPU and GPU. For 256 tokens, the average FPGA power consumption achieves a 4.72x reduction in the average power consumption of the CPU, and a 14.10x reduction in the average power consumption of the GPU. For 1024 tokens, the FPGA achieves a 14.51x reduction of the power consumption of the GPU, reaching a maximum of only 12 watts.

To calculate the total energy consumption, we need the duration of inference; therefore we introduce a new metric, the total energy consumption per token, calculated by using the inference latency and average power consumption. We measure the energy consumption per token in milliwatt hours per token.

\begin{table} [!h]

\begin{center}
\begin{small}
\begin{sc}
\caption{Total energy consumption (milliwatt hour per token, mWh/tok)}
\label{total-energy-consumption}
\vskip 0.15in
\begin{tabular}{lrr}
\toprule
 Hardware  & 256 tokens $\downarrow$  & 1024 tokens $\downarrow$ \\ 
\midrule
 CPU & 0.51 mWh/tok & 0.60 mWh/tok\\ 
 GPU & 0.33 mWh/tok & 0.34 mWh/tok\\ 
 FPGA & 0.04 mWh/tok & 0.04 mWh/tok \\
\bottomrule
\end{tabular}
\end{sc}
\end{small}
\end{center}
\end{table}

For 256 tokens, the FPGA reaches a 12.75x reduction in energy consumption over the CPU and 8.25x reduction in energy consumption over the GPU, while for 1024 tokens, the FPGA achieves a 15x reduction over the CPU and a 8.5x reduction over the GPU. Through HLSTransform, we are able to achieve high savings in energy per token.

\section{Limitations and Future Work}
We note several limitations regarding our work, and we provide potential research directions:

\subsection{Model Size}
A key limitation of our work is the on-chip memory bottlenecks that accompany FPGAs; for example, one of Xilinx’s high-end commercial FPGAs, the Virtex UltraScale+ VU19P, has an on-chip memory capacity of 224 MB \cite{ultra}. In contrast, most LLMs are much larger than the maximum size FPGAs can load on chip; for instance, Llama 2 has three pretrained LLMs of size 7, 13, and 70 billion, while GPT-3 uses 175 billion parameters \cite{llama2} \cite{gpt3}.  Since the parameters cannot be pre-initialized on on-chip memory banks due to memory constraints, the weights are instead on off-chip global memory interfaced via the AXI4 protocol, making it possible to run inference on larger models. However, external memory accesses quickly become a major bottleneck in inference latency as only 64 8-bit integers can be read per cycle. 

As a result, we limit our model size to 110M parameters. Despite the model size, there are many practical applications of similar model sizes. For instance, BERT has a base model size of 110M parameters, while ALBERT xlarge has a model size of 68M parameters; these models achieve state-of-the-art or near state-of-the-art performances on a multitude of NLP tasks and are in widespread use \cite{bert}. Several Llama variants, such as LiteLlama and TinyLlama, also have considerably smaller parameter sizes of 460M parameters and 1.1B parameters respectively, while achieving considerable generation capabilities for the size \cite{litellama} \cite{tinyllama}.

Several future directions to be explored for fitting larger models on FPGA include using greater levels of quantization (i.e. 4-bit precision) or using multiple FPGAs in unison. “Q4\_0” quantization utilizes the same quantization technique applied to 4-bit integers, and has seen success in implementations in Gerganov’s GGML library, and ongoing research exists for other quantization schemes, such as 2-bit LLMs \cite{chee}. Fully-integer quantization methods also serve as a potential research path, which both reduces parameter size and inference latency by making all weights and all calculations involve only integers, such as the ones explored in I-BERT \cite{ibert}. Model parallelism schema utilizing multiple FPGAs may also help run larger models by sharding a model across multiple FPGAs.

\subsection{Batch Size}
Another limitation of our work is our focus on the non-batched inference domain; i.e. inference with batch size 1. The large VRAM capacity and parallel computation nature of GPUs make the GPUs suitable for tasks requiring high throughput, which may make the GPU overall more power efficient in the high batch regime. An interesting future research direction is the optimization of batched inference on FPGAs.

\section{Conclusion}

We propose a new hardware accelerator for transformers on FPGA, HLSTranform, which achieves up to a 12.75x reduction and 8.25x reduction in total energy consumption per token, compared to a 2.3 GHz Intel Xeon Broadwell E5-2686 v4 CPU and a NVIDIA RTX 3090 GPU, respectively. Our FPGA accelerator maintains 0.53x the inference speed of an RTX 3090 GPU and is 2.46x as fast as the inference speed of the Intel Xeon Broadwell E5-2686 v4 CPU; these results are achieved via synthesis combined with pipelining, memory unrolling, and memory partitioning and transfer optimizations, with the addition of 8-bit integer quantization. Through our study, we provide a proof-of-concept for the usage of High Level Synthesis (HLS) as a much quicker way of prototyping FPGA designs.

As transformers become more widely used and as model sizes continue to increase, energy consumption from AI-related applications will increase correspondingly. Increased energy consumption comes with vast environmental concerns and monetary costs, as well as limiting applications that restrict power consumption such as edge computing; as a result, energy-efficient methods for inference that provide more sustainable solutions may become a much more pressing issue. We hope that our work serves as a step forward in energy-efficient methods for AI.


\bibliography{references}
\bibliographystyle{icml2023}

\newpage
\appendix
\onecolumn
\section{Appendix}

\subsection{Experimental Setup}

For all our experiments, we use a sampling temperature of 1, an empty prompt (prompt is ``''), and top-p sampling at 1. We run all our experiments 100 times and take the average for our results.

We use Karpathy's provided 110M model, which has an embedding dim of 768, 12 layers. 12 heads, 12 KV heads, and a max context length of 1024.

\subsection{Timing Results}

\begin{sc}
    
\begin{table}[!ht]
    \centering
    \caption{We obtain our timing results from the synthesis as shown below.}
    \tiny
    \begin{tabular}{|l|l|l|l|l|l|l|l|}
    \hline
        Module Name & Start Interval & Best (cycles) & Avg (cycles) & Worst (cycles) & Best (absolute) & Avg (absolute) & Worst (absolute) \\ \hline
        forward\_Pipeline\_1 & 771 & 771 & 771 & 771 & 3.084 us & 3.084 us & 3.084 us \\ \hline
        rmsnorm\_768\_Pipeline\_1 & 770 & 770 & 770 & 770 & 3.080 us & 3.080 us & 3.080 us \\ \hline
        rmsnorm\_768\_Pipeline\_2 & 771 & 771 & 771 & 771 & 3.084 us & 3.084 us & 3.084 us \\ \hline
        rmsnorm\_768\_Pipeline\_sum\_of\_squares & 5413 & 5413 & 5413 & 5413 & 21.652 us & 21.652 us & 21.652 us \\ \hline
        rmsnorm\_768\_Pipeline\_norm\_and\_scale & 23 & 23 & 23 & 23 & 92.000 ns & 92.000 ns & 92.000 ns \\ \hline
        rmsnorm\_768\_Pipeline\_5 & 770 & 770 & 770 & 770 & 3.080 us & 3.080 us & 3.080 us \\ \hline
        rmsnorm\_768\_s & 7822 & 7822 & 7822 & 7822 & 31.288 us & 31.288 us & 31.288 us \\ \hline
        round & 1 & 1 & 1 & 1 & 4.000 ns & 4.000 ns & 4.000 ns \\ \hline
        p\_hls\_fptosi\_float\_i8 & 1 & 1 & 1 & 1 & 4.000 ns & 4.000 ns & 4.000 ns \\ \hline
        quantize\_768\_Pipeline\_main\_loop & 198 & 198 & 198 & 198 & 0.792 us & 0.792 us & 0.792 us \\ \hline
        quantize\_768\_Pipeline\_2 & 770 & 770 & 770 & 770 & 3.080 us & 3.080 us & 3.080 us \\ \hline
        quantize\_768\_Pipeline\_3 & 14 & 14 & 14 & 14 & 56.000 ns & 56.000 ns & 56.000 ns \\ \hline
        quantize\_768\_s & 971 & 971 & 971 & 971 & 3.884 us & 3.884 us & 3.884 us \\ \hline
        matmul\_768\_768\_Pipeline\_x\_buff & 50 & 50 & 50 & 50 & 0.200 us & 0.200 us & 0.200 us \\ \hline
        matmul\_768\_768\_Pipeline\_xs\_buff & 5 & 5 & 5 & 5 & 20.000 ns & 20.000 ns & 20.000 ns \\ \hline
        matmul\_768\_768\_Pipeline\_VITIS\_LOOP\_225\_1 & 20900 & 20900 & 20900 & 20900 & 83.600 us & 83.600 us & 83.600 us \\ \hline
        matmul\_768\_768\_s & 20977 & 20977 & 20977 & 20977 & 83.908 us & 83.908 us & 83.908 us \\ \hline
        pow\_generic\_float\_s & 1 & 15 & 15 & 15 & 60.000 ns & 60.000 ns & 60.000 ns \\ \hline
        sin\_or\_cos\_float\_s & 1 & 18 & 18 & 18 & 72.000 ns & 72.000 ns & 72.000 ns \\ \hline
        forward\_Pipeline\_rotation1 & 119 & 119 & 119 & 119 & 0.476 us & 0.476 us & 0.476 us \\ \hline
        forward\_Pipeline\_3 & 839 & 839 & 839 & 839 & 3.356 us & 3.356 us & 3.356 us \\ \hline
        forward\_Pipeline\_4 & 839 & 839 & 839 & 839 & 3.356 us & 3.356 us & 3.356 us \\ \hline
        forward\_Pipeline\_iterate & 530 \~ 1554 & 530 & 1042 & 1554 & 2.120 us & 4.168 us & 6.216 us \\ \hline
        forward\_Pipeline\_max & 2 \~ 261 & 2 & 133 & 261 & 8.000 ns & 0.532 us & 1.044 us \\ \hline
        forward\_Pipeline\_exp & 24 \~ 56 & 24 & 40 & 56 & 96.000 ns & 0.160 us & 0.224 us \\ \hline
        forward\_Pipeline\_sum & 10 \~ 1546 & 10 & 778 & 1546 & 40.000 ns & 3.112 us & 6.184 us \\ \hline
        forward\_Pipeline\_norm & 9 \~ 25 & 9 & 17 & 25 & 36.000 ns & 68.000 ns & 0.100 us \\ \hline
        forward\_Pipeline\_10 & 66 & 66 & 66 & 66 & 0.264 us & 0.264 us & 0.264 us \\ \hline
        forward\_Pipeline\_acc & 89 \~ 1625 & 89 & 857 & 1625 & 0.356 us & 3.428 us & 6.500 us \\ \hline
        forward\_Pipeline\_residual & 61 & 61 & 61 & 61 & 0.244 us & 0.244 us & 0.244 us \\ \hline
        matmul\_768\_2048\_Pipeline\_x\_buff & 50 & 50 & 50 & 50 & 0.200 us & 0.200 us & 0.200 us \\ \hline
        matmul\_768\_2048\_Pipeline\_xs\_buff & 5 & 5 & 5 & 5 & 20.000 ns & 20.000 ns & 20.000 ns \\ \hline
        matmul\_768\_2048\_Pipeline\_VITIS\_LOOP\_225\_1 & 55460 & 55460 & 55460 & 55460 & 0.222 ms & 0.222 ms & 0.222 ms \\ \hline
        matmul\_768\_2048\_s & 55537 & 55537 & 55537 & 55537 & 0.222 ms & 0.222 ms & 0.222 ms \\ \hline
        forward\_Pipeline\_swi\_glu & 552 & 552 & 552 & 552 & 2.208 us & 2.208 us & 2.208 us \\ \hline
        forward\_Pipeline\_14 & 2050 & 2050 & 2050 & 2050 & 8.200 us & 8.200 us & 8.200 us \\ \hline
        quantize\_2048\_Pipeline\_main\_loop & 221 & 221 & 221 & 221 & 0.884 us & 0.884 us & 0.884 us \\ \hline
        quantize\_2048\_Pipeline\_2 & 2050 & 2050 & 2050 & 2050 & 8.200 us & 8.200 us & 8.200 us \\ \hline
        quantize\_2048\_Pipeline\_3 & 34 & 34 & 34 & 34 & 0.136 us & 0.136 us & 0.136 us \\ \hline
        quantize\_2048\_s & 2274 & 2274 & 2274 & 2274 & 9.096 us & 9.096 us & 9.096 us \\ \hline
        matmul\_2048\_768\_Pipeline\_x\_buff & 130 & 130 & 130 & 130 & 0.520 us & 0.520 us & 0.520 us \\ \hline
        matmul\_2048\_768\_Pipeline\_xs\_buff & 10 & 10 & 10 & 10 & 40.000 ns & 40.000 ns & 40.000 ns \\ \hline
        matmul\_2048\_768\_Pipeline\_VITIS\_LOOP\_225\_1 & 52526 & 52526 & 52526 & 52526 & 0.210 ms & 0.210 ms & 0.210 ms \\ \hline
        matmul\_2048\_768\_s & 52659 & 52659 & 52659 & 52659 & 0.211 ms & 0.211 ms & 0.211 ms \\ \hline
        forward\_Pipeline\_residual2 & 58 & 58 & 58 & 58 & 0.232 us & 0.232 us & 0.232 us \\ \hline
        matmul\_768\_32000\_Pipeline\_x\_buff & 50 & 50 & 50 & 50 & 0.200 us & 0.200 us & 0.200 us \\ \hline
        matmul\_768\_32000\_Pipeline\_xs\_buff & 5 & 5 & 5 & 5 & 20.000 ns & 20.000 ns & 20.000 ns \\ \hline
        matmul\_768\_32000\_Pipeline\_VITIS\_LOOP\_225\_1 & 864190 & 864190 & 864190 & 864190 & 3.457 ms & 3.457 ms & 3.457 ms \\ \hline
        matmul\_768\_32000\_s & 864311 & 864311 & 864311 & 864311 & 3.457 ms & 3.457 ms & 3.457 ms \\ \hline
        forward & 4160108 \~ 4892636 & 4160107 & 4377403 & 4892635 & \textbf{16.640 ms} & \textbf{17.510 ms} & \textbf{19.571 ms} \\ \hline
    \end{tabular}
\end{table}
\end{sc}



\end{document}